\documentclass[fleqn,usenatbib]{mnras}

\usepackage{newtxtext,newtxmath}

\usepackage[T1]{fontenc}
\usepackage{ae,aecompl}

\usepackage{graphicx}	
\usepackage{amsmath}	
\usepackage{amssymb}	
\usepackage{epsfig}		
\usepackage{upgreek}	

\title[Optical and X-ray absorbers in AGNs]{The relationship between X-ray and optical absorbers in active galactic nuclei}
\author[G. W. Jaffarian \& C. M. Gaskell]{Gunnar W. Jaffarian\thanks{E-mail:
gjaffari@ucsc.edu} and C. Martin Gaskell\thanks{E-mail:
mgaskell@ucsc.edu}\\
\\Department of Astronomy and Astrophysics, University of California, Santa Cruz, CA 95064
}

\begin{document}

\date{}

\pagerange{\pageref{firstpage}--\pageref{lastpage}}
\pubyear{2019}

\maketitle
\label{firstpage}

\begin{abstract}
\\
We present a large compilation of reddening estimates from broad-line Balmer decrements for active galactic nuclei (AGNs) with measured X-ray column densities. The median reddening is $E(B-V) \thickapprox 0.77 \pm 0.10$ for type-1 {to type-1.9} AGNs with reported column densities. This is notably higher than the median reddening of AGNs from the SDSS.  We attribute this to the selection bias of the SDSS towards blue AGNs. For other AGNs we find evidence of a publication bias against reporting low column densities. We find a significant correlation between $N_H$ and $E(B-V)$ but with a large scatter of $\pm 1$ dex.  On average the X-ray columns are consistent with those predicted from $E(B-V)$ for a solar neighbourhood dust-to-gas ratio.   We argue that the large scatter of column densities and reddenings can be explained by X-ray  column-density variability.  For AGNs with detectable broad-line regions (BLRs) that have undergone significant changes of Seyfert type (``changing-look" AGNs) we do not find any statistically significant differences in $N_H$ or $E(B-V)$ {compared to standard type-1 to type-1.9 AGNs}.  There is no evidence for any type-1 AGNs being Compton-thick. We also analyze type-2 AGNs and find no significant correlation between $N_H$ and narrow-line region reddening.  We find no evidence for a previously claimed anti-correlation. The median column density of LINERs is $22.68 \pm 0.75$ compared to a column density of $22.90 \pm 0.28$ for type-2 AGNs. We find the majority of low column density type-2 AGNs are LINERs, but $N_H$ is probably underestimated because of scattered light. 
\end{abstract}

\begin{keywords}
 galaxies: active -- galaxies: nuclei -- galaxies: Seyfert -- dust, extinction
\end{keywords}

\section{Introduction}

A large fraction of active galactic nuclei (AGNs) show X-ray photoelectric absorption edges from which an equivalent hydrogen absorption column density, $N_H$, can be estimated. Reported column densities range from $< 10^{20}$ cm$^{-2}$ to $> 10^{24}$ cm$^{-2}$ (i.e., Compton thick). The nature and location of the absorbing material has been long debated. Optical spectrophotometry reveals that the line-emitting regions and continua of most AGNs are significantly reddened, with colour excesses, $E(B-V)$, of the order of 0.2 or much greater (see \citealt{Gaskell17} for a review and references). Wherever there is dust there is also gas.  Observations of the Milky Way give $N_H = 5.8 \times 10^{21} E(B-V)$ cm$^{-2}$ \citep{Bohlin78}. For this standard Milky Way dust-to-gas ratio, we would therefore expect that reddenings should range from $E(B-V) \thickapprox 0.01$ (comparable to the reddenings at high Galactic latitudes in the solar neighbourhood) to $E(B-V) \gg 10$ for Compton-thick AGNs. Although there are many AGNs that are completely obscured in the optical (type-2 AGNs), there are also AGNs with very high column densities that do {\em not} show high reddenings in the optical (compare, for example the column densities given by \citealt{Risaliti+00} and \citealt{Shu+07} for Mrk 266 and NGC 3982 respectively with the H$\upalpha$/H$\upbeta$ ratios given by \citealt{SDSS12}.) This situation has been commonly interpreted as being due to relatively dust-free X-ray absorbing gas. On the other hand, when there {\em is} significant reddening there {\em must} be gas present.

In the simplest unified models (see \citealt{Antonucci93}) Seyfert 2 galaxies are thermal AGNs seen at high inclination so that the surrounding dust blocks our view of the broad-line region (BLR) and accretion disk. There has therefore been much interest in the difference between X-ray absorbing properties of type-2 AGNs and type-1 AGNs.  The na{\"i}ve predictions are that type-2 AGNs should show high column densities and that type-1 AGNs should show low column densities. There is good evidence for this (see    Goodrich et al. 1994; 
\citealt{Veilleux+97,Shi+06,Burtscher+16}).  
Another prediction is that among type-1 to type 1.9 AGNs, $N_H$ should be correlated with $E(B-V)$

By definition \citep{Khachikian+Weedman71}, the optical spectra of type-2 AGNs only show lines from the narrow-line region (NLR). For the NLR the intensity ratio of H$\upalpha$ to H$\upbeta$, which we will simply refer to as the Balmer decrement, is expected to be the Case B value ($\thickapprox 3$; see \citealt{Osterbrock+Ferland06}). This is supported by observations of other line ratios (see \citealt{Gaskell84} and \citealt{Wysota+Gaskell88}). The reddening of the NLR can therefore readily be estimated from the observed Balmer decrement.  

The studies of \citet{Dong+08} and \cite{Gaskell17} show that the intrinsic Balmer decrements of the higher density broad-line region (BLR) gas also have a Case B ratio and hence the reddening of the BLR can also be estimated from the Balmer decrement. \citet{Gaskell17} gets an intrinsic H$\upalpha$/H$\upbeta$ ratio of 2.7\footnote{The Case B ratio for the BLR is expected to be higher than for the NLR because the density is higher. For the same ionization parameter, NLR gas has a lower temperature, and hence a higher Case B ratio, because collisional-line cooling results in a lower temperature. See \citet{Osterbrock+Ferland06}. }$^{, }$\footnote{ {\citet{Gaskell17} explains that the slightly higher H$\upalpha$/H$\upbeta$ \citet{Dong+08} find for radio-loud AGNs and AGNs with double-peaked Balmer lines is a consequence of the higher reddening of AGNs seen at higher inclinations, rather 
than fundamentally different physical conditions in these AGNs}}.   \citet{Malizia+97} studied the relationship between $N_H$ and Balmer decrement for AGNs of all types (BLR decrements for types 1 - 1.9 and NLR decrements for type 2).  Contrary to the na{\"i}ve prediction, they did not find significant correlations between the NLR or BLR Balmer decrements and $N_H$. \citet{Guainazzi+01} studied the relationship between $N_H$ and the NLR Balmer decrement in just type-2 AGNs.  Contrary to predictions, they surprisingly report an {\em anti}-correlation between $N_H$ and NLR Balmer decrement (i.e., steeper Balmer decrements corresponding to {\em lower} column densities).  A physical cause of this is hard to understand.

In this paper we therefore present a compendium of column densities and Balmer decrements for a substantially larger sample of type-1 {to type-1.9} AGNs and discuss the relationship between reddening and column density and its significance. We also re-examine the relationship between column density and reddening for type-2 AGNs.

\section{Data}

In Table \ref{Table1} we present a compendium of AGN hydrogen column densities, Balmer decrements, and classifications taken from the literature. The criterion for inclusion was the availability of {\em both} an estimate of the hydrogen column density (including upper limits) from X-ray observations {\em and} a Balmer decrement.  The data are necessarily inhomogeneous and subject to publication bias.  Estimation of column densities is  model-dependent (see, for example, \citealt{Immler+03}).
Furthermore, as will be discussed below, column-density variability is very common \citep{Reichert+85}. We therefore do not quote errors for the column density even when they are estimated in the original sources. \citet{Malizia+97} found column-density variability in 70\% of the sources they could analyze. Where there were multiple values of the column density available we simply give a geometric average weighted by the number of observations (but not by any error bars).  Table \ref{Table1} also gives BLR Balmer decrements for type 1 to 1.9 AGNs and NLR Balmer decrements for type-2 AGNs and LINERS.  For these estimates we favored references with the highest quality spectra. The majority of studies have not subtracted out host-galaxy starlight from spectra before calculating Balmer decrements. This results in a systematic overestimate of the Balmer decrement when H$\upbeta$ emission is weak, especially when there is a young stellar population present with strong Balmer absorption lines. This does not affect our analysis.

In column (7) we give reddening estimates for the Balmer decrement. We have calculated $E(B-V)$ from broad-line Balmer decrements using
\begin{equation}
~~~~~~~~~~~~~E(B-V)=2.5\times1.3\times\log_{10}\left(\frac{\mbox{H}\upalpha/\mbox{H}\upbeta}{2.7}\right),
\end{equation}
where the intrinsic H$\upalpha$/H$\upbeta$ ratio for the BLR has been taken to be 2.7 following \citet{Gaskell17}. The intrinsic H$\upalpha$/H$\upbeta$ ratio has been taken to be 3.1 for the NLR. The ratio of selective extinction between H$\upalpha$ and H$\upbeta$ compared to $E(B-V)$ has been taken to be 1.3 using the Milky Way reddening curve of \citet{Weingartner+Draine01}.

Table \ref{Table1} contains 89 {type-1 to type-1.9} AGNs, and 35 type-2 AGNs or LINERS (indicated by a ``3" for their type), with their Balmer decrements and X-ray column densities. References to the sources of the measurements are given in the table.  The organization of the table is as follows.  AGNs are grouped by optical classification in order of increasing optical type. ``Changing-look" AGNs, whose optical spectra have changed enough to change the Seyfert classification are placed at the end, followed by AGNs with only upper or lower limits on the hydrogen column density. These things are noted in column (8).
All classifications for the galaxies come from the NASA/IPAC Extragalactic Database (NED). When there have been multiple type claims for the galaxies, we chose the most recent classification from NED's references. In some cases, if a claim that an object was type-2 seemed unreasonable because of obvious broad Balmer lines in the spectra, we used an older classification that was more consistent with a non-type-2 classification.

\clearpage
\begin{table*}
	\centering
	\caption{Column densities and reddenings}
	\label{Table1}
	\begin{tabular}{p{1.5in}llp{0.77in}lp{0.75in}lp{0.75in}}
		\hline
		 (1) Object Name                                                     & (2) Class & (3) H$\upalpha$/H$\upbeta$ & (4) H$\upalpha$/H$\upbeta$ Ref     & (5) log($N_H$) & (6) $N_H$ Ref                     & (7) E(B-V) & (8) Notes         \\ \hline

3C 111                                                          & 1   & 15.00 & 44*            & 22.80 & 10, 11, 73(2)                                                                 & 2.42  &               \\
3C 120                                                          & 1   & 6.46  & 1, 2          & 21.39 & 10, 76                                                                        & 1.23  &               \\
3C 445                                                          & 1   & 9.50  & 3              & 23.16 & 1, 12, 76                                                                     & 1.78  &               \\
ESO 141-G055                                                    & 1   & 4.00  & 8,9            & 21.66 & 1, 76-                                                                    & 0.55  &               \\
Fairall 51                                                      & 1   & 3.32  & 10,9           & 21.90 & 13, 76-                                                                   & 0.29  &               \\
H 1846-786                                                      & 1   & 2.40  & 41*            & 22.80 & 1                                                                             & -0.17 &               \\
I Zw 1                                                          & 1   & 4.66  & 1, 2          & 20.81 & 2                                                                             & 0.77  &               \\
IC 5063                                                         & 1   & 5.53  & 11             & 23.48 & 12, 76                                                                        & 1.01  &               \\
Mrk 1044                                                        & 1   & 2.40  & 41*            & 20.62 & 3                                                                             & -0.17 &               \\
Mrk 110                                                         & 1   & 4.22  & 1, 2, 7*      & 20.17 & 3, 76-                                                                    & 0.63  &               \\
Mrk 1239                                                        & 1   & 3.50  & 41*            & 20.93 & 3                                                                             & 0.37  &               \\
Mrk 1310                                                        & 1   & 3.97  & 7*             & 20.48 & 3                                                                             & 0.55  &               \\
Mrk 142                                                         & 1   & 2.90  & 1, 2, 7*      & 20.40 & 3                                                                             & 0.10  &               \\
Mrk 205                                                         & 1   & 4.28  & 13, 2          & 20.35 & 3, 76-                                                                     & 0.65  &               \\
Mrk 279                                                         & 1   & 4.90  & 12             & 22.90 & 5, 73, 76                                                                     & 0.84  &               \\
Mrk 3                                                           & 1   & 6.61  & 11             & 23.92 & 14, 3, 76                                                                     & 1.26  &               \\
Mrk 304                                                         & 1   & 2.86  & 1, 2, 14      & 20.16 & 16                                                                            & 0.08  &               \\
Mrk 40                                                          & 1   & 2.65  & 1, 2, 7*      & 20.36 & 3, 76                                                                         & -0.02 &               \\
Mrk 42                                                          & 1   & 3.63  & 5, 2, 7*       & 19.90 & 3                                                                             & 0.42  &               \\
Mrk 474                                                         & 1   & 3.09  & 6, 2          & 20.49 & 3                                                                             & 0.19  &               \\
Mrk 478                                                         & 1   & 4.32  & 1, 2          & 20.30 & 2                                                                             & 0.66  &               \\
Mrk 493                                                         & 1   & 2.71  & 7*             & 20.47 & 3                                                                             & 0.00  &               \\
Mrk 533 \footnotesize{2}                                        & 1   & 5.00  & 11             & 24.00 & 17                                                                            & 0.87  &               \\
Mrk 79                                                          & 1   & 5.53  & 12             & 20.97 & 1, 15, 3, 76-                                                                 & 1.01  &               \\
Mrk 876                                                         & 1   & 5.20  & 15, 6          & 19.63 & 18                                                                            & 0.93  &               \\
NGC 2110                                                        & 1   & 8.13  & 11             & 22.48 & 1, 10, 11, 5, 76                                                              & 1.56  &               \\
NGC 4051                                                        & 1   & 3.50  & 11, 6, 2       & 22.32 & 11, 22, 23, 24, 3, 73(2), 76-                                               & 0.37  &               \\
NGC 4507                                                        & 1   & 5.02  & 11             & 23.68 & 12, 25, 76                                                                    & 0.88  &               \\
NGC 4593                                                        & 1   & 2.61  & 17, 18         & 21.23 & 1, 5, 76-                                                                     & -0.05 &               \\
NGC 985                                                         & 1   & 4.67  & 6              & 20.77 & 3, 76                                                                         & 0.77  &               \\
PG 1001+054                                                     & 1   & 3.20  & 7*             & 19.37 & 30                                                                            & 0.24  &               \\
PG 1244+026                                                     & 1   & 2.95  & 7*             & 19.49 & 30                                                                            & 0.12  &               \\
PG 1448+273                                                     & 1   & 3.40  & 7*             & 19.64 & 31                                                                            & 0.33  &               \\
{[}HB89{]} 0241+622                                             & 1.2 & 16.22 & 19             & 21.91 & 12, 76                                                                        & 2.53  &               \\
Fairall 9                                                       & 1.2 & 2.59  & 10, 8          & 20.30 & 1, 76-                                                                        & -0.06 &               \\
IC 4329A                                                        & 1.2 & 11.33 & 11, 20         & 21.43 & 1, 10, 11, 35, 76                                                             & 2.02  &               \\
Mrk 335                                                         & 1.2 & 2.60  & 1, 2           & 20.55 & 1, 76                                                                         & -0.05 &               \\
Mrk 50                                                          & 1.2 & 3.30  & 6, 2, 7*       & 19.90 & 5, 76-                                                                    & 0.29  &               \\
Mrk 705                                                         & 1.2 & 4.40  & 7*             & 20.25 & 31, 3, 76-                                                                & 0.69  &               \\
NGC 4235                                                        & 1.2 & 10.54 & 25,  7*        & 21.29 &3, 76                                                                          & 1.92  &               \\
PG 1302-102                                                     & 1.2 & 2.00  & 42*            & 19.43 & 29                                                                            & -0.42 &               \\
UGC 6728                                                        & 1.2 & 7.00  & 45*            & 19.74 & 37, 76-                                                                   & 1.34  &               \\
MCG -2-58-22                                                    & 1.5 & 5.80  & 7*             & 21.38 & 1, 76-                                                                    & 1.08  &               \\
MCG 8-11-11                                                     & 1.5 & 4.28  & 12, 26          & 21.17 & 1, 76                                                                         & 0.65  &               \\
MR 2251-178                                                     & 1.5 & 5.62  & 27             & 21.56 & 1, 12, 73, 76-                                                            & 1.03  &               \\
Mrk 1152                                                        & 1.5 & 3.20  & 41*            & 20.31 & 1, 3, 76-                                                                 & 0.24  &               \\
Mrk 290                                                         & 1.5 & 3.10  & 1, 2, 7*      & 21.67 & 1, 3, 76-                                                                 & 0.20  &               \\
Mrk 509                                                         & 1.5 & 2.69  & 1, 2          & 20.81 & 1, 76-                                                                    & -0.01 &               \\
Mrk 817                                                         & 1.5 & 4.00  & 12, 23, 24     & 20.06 & 38, 76-                                                                   & 0.55  &               \\
Mrk 841                                                         & 1.5 & 4.80  & 12, 23, 24     & 20.20 & 5, 76-                                                                    & 0.81  &               \\
NGC 1275                                                        & 1.5 & 10.71 & 29             & 21.80 & 22, 39, 76                                                                    & 1.94  &               \\
NGC 1566                                                        & 1.5 & 4.18  & 30, 31         & 19.73 & 31, 76-                                                                   & 0.62  &               \\
Mrk 1218                                                        & 1.8 & 9.04  & 7*             & 20.65 & 3                                                                             & 1.71  &               \\
Mrk 334                                                         & 1.8 & 4.85  & 39             & 20.64 & 47                                                                            & 0.83  &               \\
Mrk 516                                                         & 1.8 & 30.00 & 39             & 21.59 & 3                                                                             & 3.40  &               \\
Mrk 609                                                         & 1.8 & 4.95  & 7*             & 20.77 & 3                                                                             & 0.86  &               \\
Mrk 744                                                         & 1.8 & 5.47  & 39             & 22.07 & 3, 76                                                                         & 1.00  &               \\
NGC 3660                                                        & 1.8 & 7.50  & 41*            & 20.26 & 58                                                                            & 1.44  &               \\
NGC 4395                                                        & 1.8 & 4.20  & 7*             & 21.51 & 11, 5, 76                                                                     & 0.62  &               \\
Mrk 883                                                         & 1.9 & 4.43  & 7*             & 21.15 & 3                                                                             & 0.70  &              
\\ \hline
	\end{tabular}
\end{table*}
\newpage
\clearpage
\begin{table*}
	\setcounter{table}{0}
	\caption{Continued}
	\begin{tabular}{p{1.5in}llp{0.77in}lp{0.75in}lp{0.75in}}
		\hline
		(1) Object Name                                                     & (2) Class & (3) H$\upalpha$/H$\upbeta$ & (4) H$\upalpha$/H$\upbeta$ Ref     & (5) log($N_H$) & (6) $N_H$ Ref                     & (7) E(B-V) & (8) Notes         \\ \hline
NGC 4138                                                        & 1.9 & 6.50  & 46*            & 22.63 & 11, 76                                                                        & 1.24  &               \\
NGC 4258                                                        & 1.9 & 10.00 & 7**            & 22.81 & 49, 50, 11, 3, 76                                                             & 1.85  &               \\
NGC 4388                                                        & 1.9 & 5.85  & 11, 7*         & 23.12 & 51, 11, 5, 3, 76                                                              & 1.09  &               \\
NGC 4579                                                        & 1.9 & 1.40  & 43*            & 19.30 & 11                                                                            & -0.93 &               \\
NGC 4594                                                        & 1.9 & 20.00 & 7*              & 20.40 & 3                                                                             & 2.83  &               \\
NGC 5252                                                        & 1.9 & 4.08  & 35, 7*        & 22.45 & 9, 76                                                                         & 0.58  &               \\
NGC 526A                                                        & 1.9 & 3.00  & 11             & 22.19 & 48(8), 76                                                                     & 0.15  &               \\
NGC 5273                                                        & 1.9 & 4.24  & 7*             & 20.95 & 11                                                                            & 0.64  &               \\
NGC 5506                                                        & 1.9 & 7.22  & 11, 7*         & 23.94 & 1, 10, 11, 52, 48, 73, 75, 76                                                 & 1.39  &               \\
NGC 7314                                                        & 1.9 & 20.00 & 11             & 21.72 & 1, 11, 5, 76                                                                  & 2.83  &               \\
Cen A                                                           & 2   & 3.85  & 34             & 23.06 & 12, 76                                                                        & 0.31  &               \\
Cygnus A                                                        & 2   & 8.50  & 42*            & 23.49 & 12, 76                                                                        & 1.42  &               \\
ESO 103-G35                                                     & 2   & 12.06 & 11             & 23.17 & 26, 1, 12, 73+, 76                                                         & 1.92  &               \\
F01475-0740                                                     & 2   & 5.72  & 41*            & 21.59 & 58                                                                            & 0.86  &               \\
IRAS 04575-7537                                                 & 2   & 4.53  & 36             & 22.33 & 52, 76                                                                        & 0.54  &               \\
IRAS 18325-5926                                                 & 2   & 9.91  & 37             & 22.12 & 12, 76                                                                        & 1.64  &               \\
MCG 5-23-16                                                     & 2   & 7.24  & 11, 7*         & 22.12 & 1, 10, 11, 53                                                                 & 1.20  &               \\
Mrk 270                                                         & 2   & 3.78  & 39             & 23.18 & 56                                                                            & 0.28  &               \\
Mrk 348                                                         & 2   & 6.02  & 11             & 23.10 & 26, 53, 76                                                                    & 0.94  &               \\
Mrk 463 E                                                       & 2   & 5.62  & 11             & 23.51 & 54                                                                            & 0.84  &               \\
Mrk 573                                                         & 2   & 4.20  & 39             & 22.32 & 9, 3                                                                          & 0.43  &               \\
Mrk 78                                                          & 2   & 6.50  & 39             & 22.76 & 55                                                                            & 1.05  &               \\
NGC 1320                                                        & 2   & 4.25  & 41*            & 23.60 & 19                                                                            & 0.45  &               \\
NGC 1667                                                        & 2   & 9.74  & 16             & 24.66 & 57, 58, 14                                                                   & 1.62  &               \\
NGC 1672                                                        & 2   & 6.99  & 16             & 21.80 & 57                                                                            & 1.15  &               \\
NGC 1808                                                        & 2   & 14.17 & 38             & 23.02 & 57, 59                                                                        & 2.15  &               \\
NGC 3281                                                        & 2   & 4.48  & 41*            & 23.63 & 5, 76                                                                         & 0.52  &               \\
NGC 5643                                                        & 2   & 5.58  & 16             & 22.92 & 12, 63, 77                                                                    & 0.83  &               \\
NGC 5929                                                        & 2   & 5.13  & 7*             & 20.71 & 3                                                                             & 0.71  &               \\
NGC 7172                                                        & 2   & 6.50  & 34             & 22.89 & 26, 1, 5, 76                                                              & 1.05  &               \\
NGC 7590                                                        & 2   & 3.65  & 41*            & 20.96 & 58                                                                            & 0.23  &               \\
Mrk 266SW                                                       & 3   & 6.88  & 7*             & 20.61 & 3                                                                             & 1.12  &               \\
NGC 2655                                                        & 3   & 4.70  & 41*            & 22.48 & 60                                                                            & 0.59  &               \\
NGC 3079                                                        & 3   & 20.00 & 7+*            & 24.08 & 58, 11, 76                                                                   & 2.63  &               \\
NGC 4278                                                        & 3   & 3.55  & 43*            & 20.83 & 3                                                                             & 0.19  &               \\
NGC 5005                                                        & 3   & 5.17  & 7*             & 22.88 & 58, 3                                                                         & 0.72  &               \\
NGC 6240                                                        & 3   & 20.50 & 46*            & 23.81 & 64, 3, 76                                                                     & 2.67  &               \\
3C 273                                                          & 1   & 3.05  & 42*            & 21.37 & 1, 76-                                                                    & 0.17  & changing look \\
Ark 120                                                         & 1   & 3.40  & 4              & 21.40 & 1, 76-                                                                    & 0.33  & changing look \\
MCG -6-30-15                                                    & 1.2 & 6.13  & 21, 22, 9, 7* & 21.96 & 1, 11, 73                                                                     & 1.16  & changing look \\
NGC 3227                                                        & 1.5 & 5.43  & 11             & 22.27 & 10, 11, 40, 3, 73(3), 76                                                      & 0.98  & changing look \\
NGC 3783                                                        & 1.5 & 3.37  & 11, 30         & 21.78 & 1, 10, 11, 42, 43, 24, 73(2), 76                                               & 0.31  & changing look \\
NGC 1365 \footnotemark[2]                                       & 1.8 & 27.00 & 40             & 23.67 & 4, 5, 72, 73(2)                                                               & 3.25  & changing look \\
NGC 2992                                                        & 2   & 7.08  & 39             & 21.71 & 7, 8, 5, 3, 74                                                                & 1.17  & changing look \\
NGC 4941                                                        & 2   & 7.50  & 41*            & 23.61 & 61, 76                                                                        & 1.25  & changing look \\
Mrk 590                                                         & 1   & 4.19  & 1, 6           & 20.54 & 1, 2, 3, 71-                                                              & 0.62  & changing look \\
NGC 7582                                                        & 1   & 8.32  & 11             & 23.69 & 26, 1, 26, 27, 76                                                             & 1.59  & changing look \\
NGC 7469                                                        & 1.2 & 5.42  & 12             & 21.06 & 1, 5, 3, 76                                                                   & 0.98  & changing look \\
NGC 3516                                                        & 1.5 & 3.47  & 11, 1, 2, 32  & 23.34 & 11, 41, 5, 70, 73, 76-                                                    & 0.35  & changing look \\
NGC 5548                                                        & 1.5 & 4.65  & 11, 12, 7*     & 21.41 & 1, 10, 11, 45(2), 76                                                          & 0.77  & changing look \\
NGC 6814                                                        & 1.5 & 3.67  & 6, 2           & 22.01 & 1, 76                                                                         & 0.43  & changing look \\
Mrk 1018                                                        & 1.9 & 6.82  & 39             & 20.34 & 3, 76-                                                                    & 1.31  & changing look \\
NGC 4151                                                        & 1.5 & 3.10  & 11             & 22.56 & 26, 1, 10, 41, 44, 24, 3, 73+, 76                                         & 0.19  & changing look \\
NGC 3982 \footnotemark[2]                                     & 1.9 & 4.50  & 7*             & 24.00 & 9                                                                             & 0.72  & lower limit   \\
\hline
	\end{tabular}
\end{table*}
\newpage
\clearpage
\begin{table*}
	\setcounter{table}{0}
	\caption{Continued}
	\begin{tabular}{p{1.5in}llp{0.77in}lp{0.75in}lp{0.75in}}
		\hline
		(1) Object Name                                                     & (2) Class & (3) H$\upalpha$/H$\upbeta$ & (4) H$\upalpha$/H$\upbeta$ Ref     & (5) log($N_H$) & (6) $N_H$ Ref                     & (7) E(B-V) & (8) Notes         \\ \hline
Mrk 266                                                         & 2   & 4.55  & 7*             & 25.00 & 4                                                                             & 0.54  & lower limit   \\
Mrk 1040                                                        & 1   & 6.39  & 6              & 21.56 & 1, 12(3), 76                                                                  & 1.21  & upper limit   \\
NGC 1068                                                        & 1   & 7.38  & 16             & 24.22 & 51, 1, 65, 22, 66, 11, 67, 3, 76+                                         & 1.42  & upper limit   \\
NGC 4639                                                        & 1   & 1.809 & 7*             & 19.00 & 11                                                                            & -0.57 & upper limit   \\ Messier 81                                                      & 1.8 & 7.60  & 34             & 21.56 & 22, 68, 11, 3                                                                 & 1.46  & upper limit   \\
NGC 5033                                                        & 1.8 & 4.67  & 7*             & 19.94 & 11, 3                                                                         & 0.77  & upper limit   \\
III Zw 2                                                        & 2   & 3.50  & 1, 2          & 21.97 & 1, 12, 76                                                                     & 0.17  & upper limit   \\
NGC 1058                                                        & 2   & 5.10  & 47*            & 20.78 & 11                                                                            & 0.70  & upper limit   \\
NGC 3185                                                        & 2   & 6.28  & 7*             & 20.30 & 11                                                                            & 1.00  & upper limit   \\
NGC 5194                                                        & 2   & 2.05  & 47*            & 24.00 & 69                                                                            & -0.58 & upper limit  \\
 \hline
\end{tabular}
\end{table*}

\begin{table*}
	\setcounter{table}{0}
	\caption{References}	
		\begin{tabular}{p{0.5\textwidth}p{0.5\textwidth}}
			H$\upalpha$/H$\upbeta$ References: (1) \citealt{Osterbrock77}; (2) \citealt{deBruyn78}; (3) \citealt{Osterbrock+76}; (4) \citealt{Kollatschny+81}; (5) \citealt{Phillips78}; (6) \citealt{Rudy84}; (7) \citealt{SDSS12}; (8) \citealt{Ward+78}; (9) \citealt{Glass+82}; (10) \citealt{Hawley78}; (11) \citealt{Mulchaey+94}; (12) \citealt{Cohen83}; (13) \citealt{Neugebauer+79}; (14) \citealt{Kunth79}; (15) \citealt{Grandi81}; (16) \citealt{Storchi-Bergmann95}; (17) \citealt{MacAlpine79}; (18) \citealt{Ward+82}; (19) \citealt{Margon78}; (20) \citealt{Wilson79}; (21) \citealt{Morris88}; (22) \citealt{Pineda+80}; (23) \citealt{Markarian77}; (24) \citealt{Denisyuk77}; (25) \citealt{Abell78}; (26) \citealt{Lacy+82}; (27) \citealt{Canizares78}; (28) \citealt{Rieke78}; (29) \citealt{Phillips83}; (30) \citealt{Osmer74}; (31) \citealt{Glass81}; (32) \citealt{McAlary79}; (33) \citealt{Tohline76}; (34) \citealt{Rebecchi+92}; (35) \citealt{Osterbrock93}; (36) \citealt{deGrijp92}; (37) \citealt{WGACAT}; (38) \citealt{Veron-Cetty86}; (39) \citealt{Dahari88}; (40) \citealt{Edmunds82}; (41) \citealt{Jones+09}; (42) \citealt{Torrealba+12}; (43) \citealt{Ho+95}; (44) \citealt{Buttiglione+09}; (45) \citealt{Zwicky}; (46) \citealt{Moustakas06}; (47) \citealt{RosalesOrtega10}; (*) Calculated from spectra from the reference; (**) Estimated from spectra from the reference
			&
			N$_H$ References\footnote: (1) \citealt{Turner89}; (2) \citealt{Boller96}; (3) \citealt{Pfefferkorn01}; (4) \citealt{Risaliti+00}; (5) \citealt{Winter+09}; (6) \citealt{Guainazzi02}; (7) \citealt{Weaver+96}; (8) \citealt{Mushotzky82}; (9) \citealt{Shu+07}; (10) \citealt{Weaver95}; (11) \citealt{Nandra94}; (12) \citealt{Malizia+97}; (13) \citealt{Bailon+08}; (14) \citealt{Bianchi+05a}; (15) \citealt{Tueller+08}; (16) \citealt{Kartje+97}; (17) \citealt{Bianchi+05b}; (18) \citealt{Schartel+96}; (19) \citealt{Balokovic+14}; (20) \citealt{Eracleous10}; (21) \citealt{Ruschel-Dutra+14}; (22) \citealt{Brinkmann94}; (23) \citealt{Mihara+94}; (24) \citealt{Crenshaw12}; (25) \citealt{Matt+04}; (26) \citealt{Warwick+93}; (27) \citealt{Rivers+15}; (28) \citealt{Nardini+15}; (29) \citealt{Rachen96}; (30) \citealt{Wang96}; (31) \citealt{Walter93}; (32) \citealt{Saez+12}; (33) \citealt{Marinucci+14}; (34) \citealt{Rozanska+04}; (35) \citealt{Madejski+95}; (36) \citealt{Miniutti+10}; (37) \citealt{Winter+08}; (38) \citealt{Winter+10}; (39) \citealt{Rhee94}; (40) \citealt{Markowitz+09}; (41) \citealt{Morse+95}; (42) \citealt{Turner+93}; (43) \citealt{George95}; (44) \citealt{Weaver+94}; (45) \citealt{Mehdipour+15}; (46) \citealt{Ursini+15}; (47) \citealt{Prieto02}; (48) \citealt{Risaliti02}; (49) \citealt{Pietsch+94}; (50) \citealt{Makishima+94}; (51) \citealt{Cappi+06}; (52) \citealt{Smith96}; (53) \citealt{Mulchaey+93}; (54) \citealt{Imanishi04}; (55) \citealt{Gilli10}; (56) \citealt{Guainazzi+05}; (57) \citealt{Awaki93}; (58) \citealt{Ying12}; (59) \citealt{Junkes+95}; (60) \citealt{Martin08}; (61) \citealt{Vasudevan+13}; (62) \citealt{Puccetti+14}; (63) \citealt{Guainazzi04}; (64) \citealt{Puccetti+16}; (65) \citealt{Marshall+93}; (66) \citealt{Ueno+94}; (67) \citealt{Bauer+15}; (68) \citealt{Petre+93}; (69) \citealt{Fukazawa+01}; (70) \citealt{Turner+11}; (71) \citealt{Rivers+12}; (72) \citealt{Risaliti+05}; (73) \citealt{Gofford+13}; (74) \citealt{Shu+10}; (75) \citealt{Sun+17}; (76) \citealt{Ricci+17}; (77) \citealt{Matt+13}; (78) \citealt{Wang+10}; (79) \citealt{Zhang+06}; (80) \citealt{Yamada+18}; (81) \citealt{Barcons+03}; (82) \citealt{Piconcelli+04}; (83) \citealt{Gallo+06}
            
            \footnotemark[2] \footnotesize{See text}  
            
			\footnotemark[3] \footnotesize{It is left unclear as to whether Galactic column densities are subtracted from the column density in most references, but the usual small adjustment for Galactic column density does not affect our analysis}

		\end{tabular}
\end{table*}

\clearpage

\section{Analysis}

 Table \ref{AGN comparison} gives the median column densities with errors in the medians and median reddenings for the various optical sub-classes.  It should be remembered that in Tables \ref{Table1} and \ref{AGN comparison} the reddenings of the type-1 to type-1.9 AGNs refer to the BLR, whilst the reddenings of the type-2 AGNs and LINERs refer to the NLR.

\subsection{Type-1s {and intermediates}}

The median reddenings of the BLRs of the type-1.8 and 1.9 AGNs in Table \ref{AGN comparison} are not unusual (see \citealt{Heard+Gaskell16}).  However, for type-1 and 1.5 AGNs the median reddenings are approximately three times the median reddening of type-1 AGNs in the SDSS calculated assuming the same intrinsic BLR Balmer decrement \citep{Gaskell17}.  This is not surprising since SDSS AGNs are heavily biased towards bluer AGNs because of the SDSS colour selection.  The type-1 to type-1.5 AGNs in Table \ref{AGN comparison} also have reddenings some 50\% higher than the reddenings \citet{Gaskell17} found for the Seyfert 1s of \cite{Osterbrock77}. The higher median $E(B-V)$ in Table \ref{AGN comparison} points to reported column densities being biased against low X-ray columns and reddenings.  To test this we looked at the reddenings of the \citet{Osterbrock77} AGNs with and without reported X-ray columns.  For those without reported X-ray columns the median reddening is $E(B-V) = 0.31 \pm 0.10$ whilst for those with reported X-ray columns $E(B-V) = 0.63 \pm 0.12$.  This supports the conclusion that there is a bias towards reporting X-rays columns for AGN with higher column  densities and higher reddenings.

\begin{table}
	\centering
	\caption{Median column densities and reddenings for different AGN types.}
	\label{AGN comparison}
	\begin{tabular}{ccc}
		\hline
		AGN type & Median log($N_H$) (cm$^{-2}$) & Median $E(B-V)$ \\
		(1) & (2) & (3) \\ \hline
		1 & $20.93 \pm 0.24$ & $0.63 \pm 0.12$ \\
		1.5 & $21.49 \pm 0.31$ & $0.59 \pm 0.15$ \\
		1.8 & $21.14 \pm 0.43$ & $1.22 \pm 0.40$ \\
		1.9 & $22.19 \pm 0.50$ & $1.09 \pm 0.35$ \\
        All 1 - 1.9 & $21.26 \pm 0.18$ & $0.77 \pm 0.10$ \\ 
        \\
		2 & $22.90 \pm 0.28$ & $0.85 \pm 0.14$ \\
		LINER & $22.68 \pm 0.75$ & $0.92 \pm 0.55$ \\
		\hline
	\end{tabular}
\end{table}

Figure 1 shows the relationship between $E(B-V)$ and $N_H$ for type-1 to type-1.9 Seyferts. This is a sample size about three times larger than that of the \citet{Malizia+97} sample. It can be seen that there is a modest correlation between column density and reddening.  The Spearman rank correlation coefficient $r=0.43$, giving a single-tailed probability $p = 1.43*10^{-5}$, for the null hypothesis of no correlation.  If we assume, following \citet{Malizia+97}, that the scatter is mostly in $N_H$ and perform a regression of $\log N_H$ on $\log E(B-V)$ we get a slope that is very similar to the linear relationship of \citet{Bohlin78}.  If we assume equal errors in $N_H$ and $\log E(B-V)$ and calculate a linear OLS bisector fit \citep{Isobe+90} it can be seen in Figure 1 that we get a steeper slope.

As an illustration of the effect of publication bias against reporting non-detections of $N_H$ we show in the Figure 2 the effects of assuming that the 18 AGNs from \citet{Osterbrock77} without published column densities had $\log N_H = 20$ (i.e., towards the lower end of the range in Figure 1). This raises the Spearman rank correlation coefficient to $r = 0.47$ lowering the single-tailed probability of the null hypothesis being correct to $p = 1.8*10^{-7}$.  As can be seen by comparing Figures 1 and 2, the effect on the slope of the linear regression of $\log N_H$ on $\log E(B-V)$ and the offset of this from the \citet{Bohlin78} relationship is small.

\begin{figure} 
	\label{NoOster}
	\begin{center}
		\includegraphics[width = 1.05\linewidth]{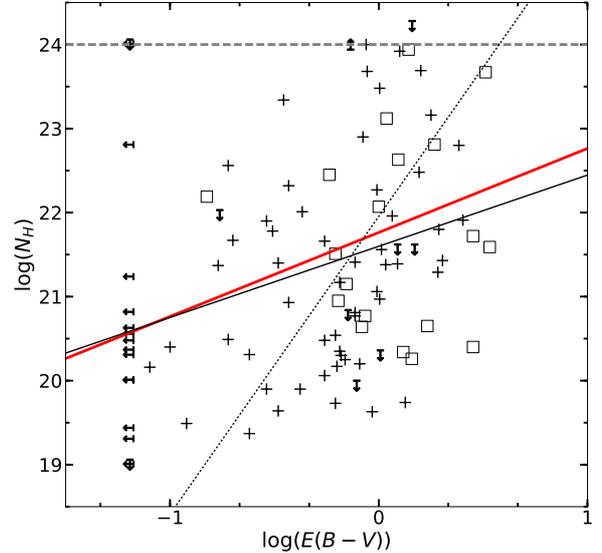}
        \caption{ X-ray column densities versus $\log E(B-V)$ for 89 type-1 and intermediate AGN. Type-1 and type-1.5 AGNS are shown as black crosses and type-1.8 and type-1.9 AGNs are shown as hollow squares. For negative reddenings in Table \ref{Table1}, $\log E(B-V)$ has been set to -1.2.  The thick red line indicates the expected relationship between column density and reddening from \citet{Bohlin78} and the dashed horizontal line indicates where an AGN becomes Compton-thick. The solid black line is a linear regression of $\log N_H$ on $\log E(B-V)$ (i.e., assuming that errors in $\log N_H$ dominate) and the dotted line is a linear OLS bisector fit \citep{Isobe+90} assuming roughly equal errors in each axis.}
	\end{center}
\end{figure}

\begin{figure} 
	\label{Oster}
	\begin{center}
		\includegraphics[width = 1.05\linewidth]{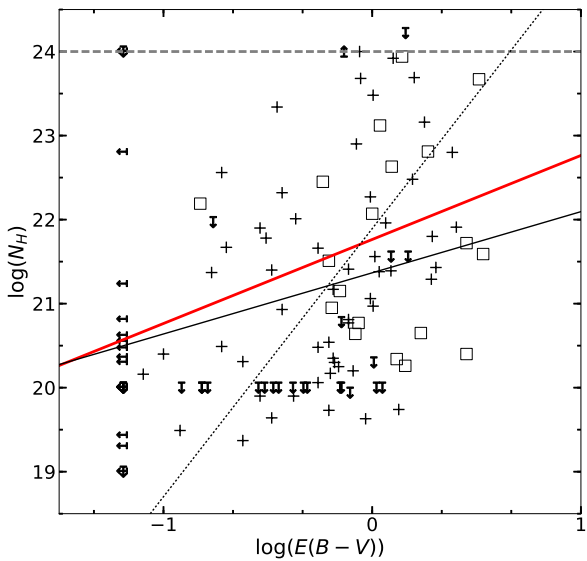}
        \caption{Same as Figure 1 except that we have included as upper limits in $\log N_H$ the 18 AGNs from \citet{Osterbrock77} that do not yet have reported hydrogen column densities.  We have assumed upper limits of $\log N_H = 20$. Note there is only a slight change in the regression lines. This indicates that publication bias does not have a strong effect on the correlations.}
	\end{center}
\end{figure}
 
\subsection{Comparing type-1 {to type-1.9} AGNs, type-2 AGNs, and LINERS}

It has long been known that type-2 AGNs have higher column densities than type-1 AGNs (e.g., \citealt{Malizia+97}) and the larger sample presented here confirms this. A t-test comparing the column densities for type-1 and intermediate AGNs to type-2 AGNs in Table \ref{Table1} gives a probability of only $p = 1.25\times 10^{-6}$ (single-tailed) for the $N_H$ distributions being the same.  

 In Table \ref{AGN comparison} the median $\log N_H$ for both the Seyfert 2s and the LINERS is an order of magnitude larger than for the type-1 {to type-1.9} AGNs. Both the X-ray column densities and the median reddenings of the NLRs of the two classes appear to be similar.  The similar reddenings implies that the amount of dust in or around the regions producing the narrow emission lines is similar in the two classes of object.  There is no evidence from the information considered here that the type-2 AGNs and LINERS are fundamentally different (see {\citealt{Ho08}} and \citealt{Antonucci12} for extensive discussion).

\subsection{The relationship between X-ray column density and NLR reddening for type-2 AGNs}

The NLR is considerably more extended than the BLR and X-ray-emitting regions of an AGN.  Comparison of NLR and BLR reddenings (see Figure 3 of \citealt{Heard+Gaskell16}) shows that the BLR reddening is generally at least the same as the NLR reddening, but frequently greater.  \citet{Heard+Gaskell16} demonstrate that the relationship between NLR and BLR reddenings and relationships between other quantities (equivalent widths and broad/narrow line ratios) are consistent with the bulk of the dust blocking the inner regions of an AGN being located between the NLR and BLR, but with more distant dust, either associated with the NLR or in the host galaxy, also contributing.

Figure 3 shows the relationship between $N_H$ and the reddening of the NLR for the type-2 AGNs and LINERs. As can be seen, type-2 AGNs show {\em no} significant correlation (Spearman rank correlation coefficient, $r = 0.16$, with a single-tailed $p = 0.18$) between column density and reddening of their NLRs.  Several things can be noted: 
\begin{enumerate}
\item The data do not support the claimed inverse correlation of \citet{Guainazzi+01} (shown by the curved line in Figure 3).  
\item{The column density is {\em higher} than predicted from the NLR by the \citet{Bohlin78} relationship in the majority of cases.}  
	\item{The data  set does not include a large number of Compton-thick AGNs.}
\end{enumerate}

\begin{figure} 
	\label{S2s}
	\begin{center}
		\includegraphics[width=1.05\linewidth]{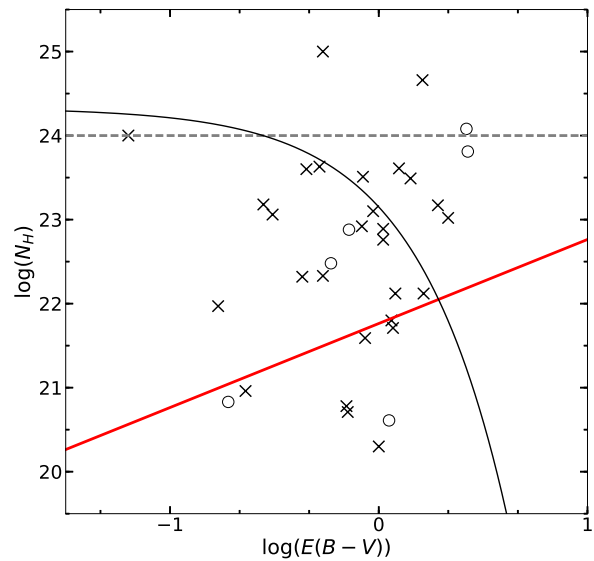}
        \caption{X-ray column densities versus NLR log $E(B-V)$ for 35 type-2 AGNs and LINERs.  Type-2 AGNs are shown as crosses and LINERs as open circles. The thick red line indicates the expected relationship between column density and reddening from \citet{Bohlin78}. The black curve shows the correlation claimed previously by \citet{Guainazzi+01}. The dashed grey line indicates where an AGN becomes Compton-thick.}
	\end{center} 
\end{figure}

\section{Discussion}

\subsection{No dependence of Seyfert 2 column densities on NLR reddening}

The lack of correlation between $N_H$ and $E(B-V)$ for type-2 AGNs is not unexpected since the NLR gas is not usually on our direct line-of-sight to the black hole in type-1 objects (see \citealt{Fischer+13}). There would only be a correlation if there were a common obscuring dusty screen covering both the NLR and the inner regions.  

The higher X-ray columns in type-2 AGNs than those predicted from the NLR reddenings are consistent with the standard model where the obscuration of type-2 AGNs is withing the NLR.

There is no support for the curious anti-correlation claimed by \citet{Guainazzi+01} between $E(B-V)$ and $N_H$ for Seyfert 2s. Such an anti-correlation would be hard to understand physically since it would imply that there was more dust when the total column density of gas was lower. The anti-correlation they reported is probably an artifact of excluding very high $N_H$ values.

\subsection{Column-density variability as the cause of the scatter in type-1 AGNs}

Although, as discussed above, there is a modest correlation between $E(B-V)$ and $N_H$ for Seyfert 1s, there a large scatter. We find this to be $\pm 1.4$ dex. \citet{Malizia+97} find significant variations in $N_H$ for most AGNs with repeated observations.  Based on their reported variability given in their Table 1, we calculate that the scatter in $\log N_H$ due to variability and measuring error for a typical observation of an AGN is $\pm 0.9$ dex. 

In addition to the effects of variability and measuring errors on $N_H$, there will also be errors in estimating $E(B-V). \, $H$\upalpha$ and H$\upbeta$ fluxes are typically measured to an accuracy of $\pm 15$\%.  Thus the error in $\log E(B-V)$ will be about 0.40 dex.   

Adding these effects in quadrature predicts a scatter of $\pm 1.0$.  Thus {\em all of the scatter in Figure 1 could be due to column density variability and measuring errors} if the reddening of the emission lines and continuum does not vary with changes in the hydrogen column along our line of sight to the X-ray emitting region.  This is reasonable since partial coverage of the X-ray-emitting inner regions of AGNs \citep{Reichert+85} implies that the size of absorbing clouds is comparable in size to the X-ray-emitting region (less than a light-day in a typical Seyfert considered here).  Variability of broad emission line profiles \citep{Gaskell+Harrington18} implies that dust clouds obscuring the BLR can often have sizes smaller than the typical BLR radii of a few light weeks.  Whilst there can be changes in reddening of the BLR, they will be less than changes in $N_H$ because a large area is being averaged.  They will also generally not occur at the same time.

AGNs can show substantial changes in their  Balmer line profiles.  Also their Balmer decrements show changes both with {\em velocity} (i.e., across line profiles) and with {\em time} (see \citealt{Gaskell+Harrington18} and references therein).  \citet{Gaskell+Harrington18} have demonstrated that these changes, and puzzling changes in the time delays in response to continuum variability, can be explained by small dusty clouds partially obscuring the BLR.  They propose that these clouds are part of the bi-conical outflow of dust in AGNs.

The timescale of variability of X-rays from the inner corona of the accretion disc shows that the X-ray-emitting region is much smaller than the lower-ionization BLR producing the Balmer lines.  If the scenario of partial obscuration of the BLR is correct, then the small compact clouds must also at times cover the inner regions producing most of the X-rays. \citet{Lamer+03} show a good example of the variation in $N_H$ caused by a compact cloud passing front of the inner region of NGC~3227.   Small-scale structure in absorbing clouds will introduce a lot of scatter in $N_H$ in Figures 1 \& 2.  Our sample in Table 1 includes galaxies with high column densities but low reddenings. These can be explained by having a compact obscuring cloud covering the X-ray corona, but mostly leaving the outer BLR unobscured.  Conversely, when the BLR is reddened, but the X-rays are unobscured, we would be seeing the innermost regions through a hole in the obscuring material.

\subsection{$N_{H}$ predicted from $E(B-V)$}

In Figure 3, $N_H$ in type-2 AGNs is systematically greater than predicted by $E(B-V)$.  This is unsurprising since there is much more central obscuration in type-2 AGNs and the line ratio is only giving the ratio from the NLR.  However, for type-1 {to type-1.9} AGNs (see Figures 1 and 2) there is no significant difference between the average observed and average predicted values if one uses the Galactic dust-to-gas ratio.  The gas along our line of sight, much of it presumably expelled by the AGN, is therefore dusty.

\subsection{Scattered light?}

In type-2 AGNs there can be a substantial contribution of scattered light in X-ray spectra. This will dilute the photoelectric edge due to the absorbed X-rays and hence lead to lower column density estimates. We suggest that this is also going on to varying degrees in type 1 - 1.9 AGNs.  The effect will be to have moved points systematically to the bottom of Figure 1. Because of the difficulties of estimating $N_H$ it would be interesting to compare high-quality spectra of type-1 AGNs with both high and low reddenings to try to understand the differences.  We suggest that a similar problem can arise in the optical.  When the extinction is substantial even a small addition of scattered light to the optical spectra will change the observed Balmer decrement.  

\subsection{``Changing-look'' AGNs}

\begin{figure} 
	\label{CL}
	\begin{center}
		\includegraphics[width = 1.05\linewidth]{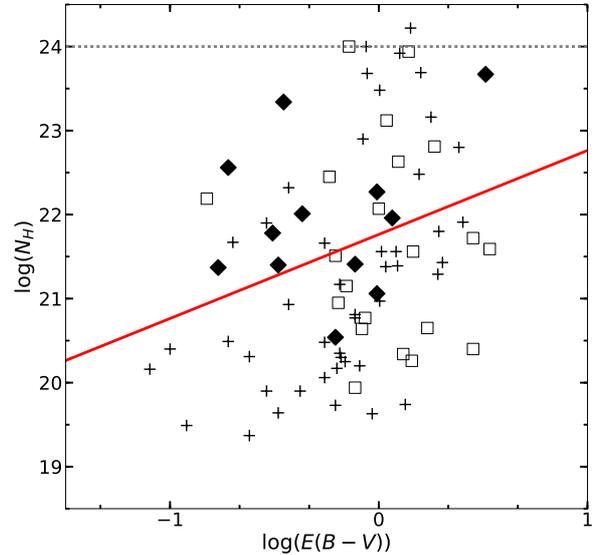}
        \caption{As for Figure 1 except that changing-look AGNs are shown as solid black diamonds and upper limit reddenings are not included.}
	\end{center}
\end{figure}

As noted above, large changes in $N_H$ are common in type-1 {and intermediate} AGNs.  There are also some AGNs that show substantial changes in their optical Seyfert type (see \citealt{Oknyansky+17}). These are known as ``changing-look'' AGNs (CL AGNs). The optical changes are in the strength of the broad components of the Balmer lines and in the continuum shape.  The number of CL AGNs known at present is small, but many of them are well-know AGNs that have long been well-studied.   This suggests that the real fraction of CL AGNs is quite high and that, as monitoring continues, many more AGNs will be classified as CL AGNs.  

In Figure 4 we show the $N_H$ and $E(B-V)$ values for known optical CL AGNs compared with other AGNs that have not (yet) been seen to change their types.  From this we see that, with the present sample, there are no obvious major differences between the CL AGNs and the AGNs not seen to have changed their types.  Whilst the median values of both $N_H$ and $E(B-V)$ are higher for type 1.8 - 1.9 AGNs compared with type 1 - 1.5 AGNs (see Table \ref{AGN comparison}), there is no statistically significant difference with the current small sample in these values for the CL AGNs.  The $N_H$ and $E(B-V)$ values in the literature thus do not at present reveal any striking differences from non-CL AGNs.  This is perhaps unsurprising since, by definition, CL AGNs are changing between types 1.8 - 1.9 and types 1.0 - 1.5.  The lack of a significant difference is consistent with the idea that whatever is causing changing-looks, it is not due to fundamentally different amounts of obscuring material in CL AGNs.

\subsection{Compton-thick type 1s?}

In the simplest unified models (such as the ``straw-person model" of \citealt{Antonucci93}), Seyfert 1 galaxies should {\em not} be Compton-thick. Alleged type-1 AGNs with claimed high column densities were therefore carefully scrutinized. We checked both whether the column density estimates are correct and whether the AGNs are really type-1s.

\subsubsection{Incorrect column densities?}

We examined the X-ray spectra of the three Seyfert 1s with published X-ray column densities around $10^{24}$ cm$^{-2}$. After re-examining the X-ray spectra of these galaxies, NGC 3982, NGC 7674 (also known as Mrk 533), and NGC 6552, we found that, contrary to previous claims, they are {\em not} Compton-thick. The problem in all three cases is that the sources are weak, the X-ray spectra are noisy, and the background subtraction is uncertain. Two of the three show negative counts at low energies, thus indicating that too much background has been subtracted. The third shows a very strange spectrum. After adjustment of the background, probable absorption edges are seen in the two with negative spectra. Thus these AGNs {\em cannot} be Compton-thick because it is impossible for a Compton-thick object to show a strong absorption edge on its X-ray spectrum. Another object, NGC 1365, has a spectrum that shows a complex shape that is hard to interpret (see \citealt{Risaliti+09}). It would be beneficial to check more of the spectra to verify the column densities.

\subsubsection{Are the classifications correct?}

Optical spectra of allegedly type-1 AGNs with very high column densities were examined to check the classification. In all cases, these AGNs were found to be type-2s or at best type-1.9s. Details are as follows:

Even though some would classify NGC 424 as a Seyfert 1, the spectrum of \citet{Fosbury+Sansom83} clearly shows that it is a Seyfert 2 (see their Figure 1). IC 5063, although classified as a Seyfert 1, can be clearly seen as a Seyfert 2 by the spectrum of \citet{Caldwell+Phillips81}. NGC 7582, although classified as a Seyfert 1, can be seen as a Seyfert 2 by the spectrum of \citet{Ward+78} Figure 5a. NGC 4507 and 3C 445 are seen as Seyfert 1s but are classified as Seyfert 1.9s by \citet{Veron+81}. NGC 1386 is misclassified as Seyfert 1, and is better classified as a Seyfert 2 by \citet{Terashima+02}.

We thus conclude from the re-examination of X-ray and optical spectra that so far there is {\em no} evidence for any type-1 AGNs being Compton-thick.

\subsection{Unobscured Type 2s?}

Among the Seyfert 2 data are some Seyfert 2s that show a dramatically low column density for the proposed models. Much of this could be because of failure to recognize the effects of scattered light, but, as just discussed for type-1 AGNs, there can also be an issue with classification. The inclusion of {non-thermal AGNs} among the Seyfert 2 sample could give misleading results since non-thermal AGNs are fundamentally different (see \citealt{Antonucci12}) and might not have the surrounding dust. We found that a number of AGNs classified as type-2 in the literature were classified elsewhere as LINERs.  These include, for example, NGC 4594 and NGC 5005 \citep{Terashima+02} as well as six other AGNs.

\section{Conclusions}

We have made a compilation of reported estimates of the hydrogen column density, $N_H$, from X-ray spectra, and have made reddening estimates from broad and narrow Balmer decrements.  The median BLR reddening is $E(B-V) \thickapprox 0.77 \pm 0.10$ for type-1 {to type-1.9} AGNs with measured X-ray column densities.  This is substantially greater than the the median reddening for AGNs found in the SDSS because the latter are biased towards blue AGNs by the SDSS selection criteria. We also find evidence of a publication bias against reporting low column densities. 

We find that the observations of type-1 {to type-1.9} AGNs are consistent with them following the standard Galactic relationship between $E(B-V)$ and $N_H$ of \citet{Bohlin78}. {There is a large scatter of $\pm 1.4$ dex (without accounting for uncertainties in measurements) but the correlation is statistically significant. The Spearman one-tailed probability for the null hypothesis of no correlation is $p = 1.4 \times 10^{-5}$.}

We find that known variability of $N_H$ is the dominant factor causing the scatter in the relationship between $E(B-V)$ and $N_H$ and that this plus observational errors in estimating $E(B-V)$ can fully account for the scatter.  

We find that all previously reported claims Seyfert 1s with high column densities are spurious either because of misclassification of the optical spectra or low-quality X-ray spectra.  We find no evidence for any cases of Compton-thick Seyfert 1s. 

We find no significant correlation between column density and reddening of the NLR in type-2 AGNs. The puzzling anti-correlation claimed by \citet{Guainazzi+01} between the reddening of the NLR and $N_H$ in type-2 AGNs is not found

The median column density of LINERs is $22.68 \pm 0.75$ compared with $22.90 \pm 0.28$ for type-2 AGNs.  We find that a number of reported Seyfert 2s with low column densities are LINERs, but $N_H$ is probably being underestimated in genuine type-2 AGNs because of scattered X-ray light diluting the photoelectric edge. 

\section*{Acknowledgments}
We are grateful to Jane Turner, Ski Antonucci, and an anonymous referee for helpful comments on the paper.

\label{lastpage}

\end{document}